%% file: main.tex
\documentclass[journal]{IEEEtran}
\usepackage{amsmath,amssymb}
\usepackage{graphicx,graphics,color,psfrag}
\usepackage{cite,balance}
\usepackage[labelsep=period]{caption}
\usepackage{framed}
\allowdisplaybreaks
\usepackage{accents}
\usepackage{amsthm}
\usepackage{bm}
\usepackage{url}
\usepackage[english]{babel}
\usepackage{multirow}
\usepackage{enumerate}
\usepackage{cases}
\usepackage{stfloats}
\usepackage{dsfont}
\usepackage{color,soul}
\usepackage{amsfonts}
\usepackage{cite,graphicx,amsmath,amssymb}
\usepackage{hhline}
\usepackage{graphicx,graphics}
\usepackage{array,color}
\usepackage{mathtools}
\usepackage{amsmath}
\usepackage[T1]{fontenc}
\usepackage{float}  
\usepackage{subcaption}
\usepackage{stfloats}
\usepackage{siunitx}
\usepackage{algorithm}
\usepackage{algpseudocode}

\def \re {{\mathrm{Re}}}

\include{header}

\begin{document}

\title{
Beam-focusing Analysis for Modular XL-arrays: Effect of Time Synchronization Errors
}
\author{
Mingjiang Wu,
Changsheng You, {\em Member, IEEE},
and
Xianfu Lei, {\em Member, IEEE}

  \thanks{
  {Copyright (c) 20xx IEEE. Personal use of this material is permitted.
However, permission to use this material for any other purposes must be
obtained from the IEEE by sending a request to pubs-permissions@ieee.org. }
}

  \thanks{
Mingjiang Wu and Changsheng You are with the Department of
Electronic and Electrical Engineering, Southern University of Science
and Technology, Shenzhen 518055, China (e-mails: wumj@sustech.edu.cn; youcs@sustech.edu.cn).

Xianfu Lei is with the School of Information Science and Technology, Southwest Jiaotong University, Chengdu 610031, China (email:xflei@swjtu.edu.cn).

\textit{(Corresponding author: Changsheng You.)}
}

}

\maketitle
\thispagestyle{empty}
\pagestyle{empty}

\begin{abstract}
For \emph{near-field} communications, it is a hardware-efficient means to form an extremely large-scale array (XL-array) by concatenating multiple modular arrays (also referred to as subarrays). In this letter, we aim to investigate the effect of \emph{time synchronization} errors among transmissions of different subarrays on the beam-focusing performance. To this end, we first characterize the beam pattern function when the transmit beamforming is designed based on maximum ratio transmission (MRT) under the premise of perfect time synchronization. As this function is highly difficult for analysis, we then consider a typical case with two subarrays. Interestingly, we show that for this case, the beam-focusing effect still persists even in
the presence of time synchronization errors, while the
focused location is deviated from the user location with an angle offset upper-bounded by $1/M$, where $M$ denotes the number of antennas in each subarray. Subsequently, for the general case with multiple subarrays, despite analytical intractability, we numerically show that time synchronization errors give rise to an \emph{imbricated} (instead of focused) beam pattern. This may significantly degrade multi-user communication performance in practice due to the reduced desired signal power and increased inter-user interference. 
\end{abstract}

\begin{IEEEkeywords}
Extremely large-scale array,  near-field communications, modular array, time synchronization. 
\end{IEEEkeywords}

\section{Introduction}

The future 6G networks are migrating to higher frequency bands (e.g., FR3) for exploiting more spectrum resources. For fixed base station (BS) sites, this trend inevitably necessitates deploying more and more antennas at the BSs to compensate for more severe path-loss. As such, the conventional far-field planar-wavefront propagation model may not be accurate enough, giving rise to the new \emph{near-field} communication scenario characterized by \emph{spherical} wavefronts. These new channel characteristics allow for focusing the beam energy around the user location/region, hence greatly enhancing the received signal power and mitigating inter-user interference. Moreover, it also enables new applications such as near-field  sensing/localization, physical-layer security, and  wireless power transfer \cite{You10858129}, \cite{Zhang10500404}.  

Existing works on near-field communications have mostly considered  \emph{collocated}  extremely large-scale arrays  ({XL-arrays}), for which all antennas are placed on a single and highly integrated platform \cite{Haiquan10496996}. Various design issues were studied for collocated XL-arrays, such as beam-focusing analysis, beam training and channel estimation, beamforming designs \cite{Mingyao9693928}, \cite{Haiyang9738442}. However, it is worth noting that collocated XL-arrays generally pose high implementation challenges in practice, due to the difficulty and complexity in fabrication and array feed network layout design. To tackle these issues, \emph{modular} arrays (also referred to as subarrays) emerge as a promising means to collectively form XL-arrays for achieving near-field beam-focusing gain, which enjoy much lower complexity in fabricating multiple low-dimensional subarrays~\cite{Li10545312},~\cite{Gon11122484}. Moreover, modular XL-arrays allow for flexible replacement of faulty subarrays, hence yielding superior maintainability. This thus inspires several recent works to study the transceiver designs of modular XL-arrays. For example, it is shown in \cite{Zhang11127210} that modular XL-arrays can provide higher spatial resolution and spectral efficiency than collocated XL-arrays. Nevertheless, these works mostly assumed perfect time synchronization among transmissions of different subarrays, thereby overlooking the impact of time synchronization errors on near-field communication performance arising from various practical factors such as imperfect clock synchronization and unequal cable lengths~\cite{Tian11359250}.

Motivated by the above, we consider in this letter a (compact) modular XL-array system and study the impact of  time synchronization errors on the beam-focusing performance. Specifically, based on the maximum ratio transmission (MRT) beamforming, we obtain  a near-field  beam pattern function accounting for time synchronization errors. Since the resulting expression is complex and difficult to analyze directly, we first examine a representative scenario involving two subarrays to gain useful insights. Interestingly, we show that even with time synchronization errors, 
the beam-focusing effect still persists, while the focused location is deviated from the user location with an angle offset upper-bounded by $1/M$, with $M$ denoting  the number of antennas in each subarray. Subsequently, we further consider the case with multiple subarrays and  reveal that time synchronization errors induce an \emph{imbricated} beam pattern, characterized by a \emph{beam-splitting} effect. Such a beam pattern not only reduces the beamforming gain for the intended user, but also induces severe inter-user interference, thereby degrading system performance in multi-user scenarios.

\section{System Model}\label{section:System Model}

\begin{figure}
\captionsetup{singlelinecheck = false, justification=justified}
   \centering
   \includegraphics[width=2.4in]{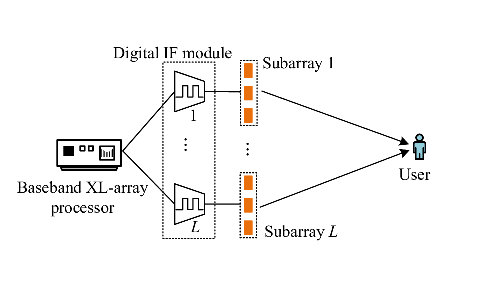}
   \vspace{-0.2in}
   \captionsetup{font={footnotesize}}
    \caption{Considered modular XL-array communication system.}
   \vspace{-0.25in}
\label{systemmodel}
\end{figure}

We consider a downlink millimeter-wave (mmWave) communication system as shown in Fig. \ref{systemmodel}, where a compact modular  antenna array  composed of  $L$ subarrays is deployed at the BS, each containing $M$ antennas, and hence, $N = LM$
antennas in total. The subarrays are connected to a baseband  processor module via digital intermediate frequency (IF) modules. 
 Moreover, we consider \emph{imperfect time synchronization} among the $L$ subarrays caused by non-ideal hardware components~\cite{Tian11359250},~\cite{Zhang11162269}.  
Without loss of generality, all subarrays are positioned along the $y$-axis and  the whole array is centered at the origin of the Cartesian coordinate. {For the compact modular array,  the inter-subarray spacing is zero, which results in half-wavelength inter-antenna spacing for all antennas. As such, the position of antenna $n$ is $ (0, \delta_{n}d)$,  where $\delta_{n} = \frac{2n-N+1}{2}$, $ \forall n \in  \mathcal{N} \triangleq \{0, 1, ..., N-1\}$, and $d = \frac{\lambda_{c}}{2}$ is the inter-antenna spacing  with $\lambda_{c}$ being the carrier wavelength.
}

 \vspace{-0.1in}
\subsection{Near-Field Channel Model }

 {We consider the case where the user is located in the near-field region of the BS, for which the distance between the user and BS is smaller than the Rayleigh distance \cite{Zhang9913211}, defined as ${2 D^2}/{\lambda_c}$ with $D$ denoting the XL-array aperture.} Let  $\mathbf{h}^H = \left[ \mathbf{h}_{1}^{H}, ...,   \mathbf{h}_{\ell}^{H}, ...,  \mathbf{h}_{L}^{H} \right] \in \mathbb{C}^{1\times N}$ denote the  channel from the BS to the user, where $\mathbf{h}_{\ell}^{H} \in \mathbb{C}^{1\times M}$, $\ell \in \mathcal{L} \triangleq \{1, 2, ..., L\}$,  is the {channel} of the $\ell$-th subarray. To characterize the  beam-focusing performance, we consider  the line-of-sight (LoS) near-field  channel model, for which the subarray channel,  $\mathbf{h}_{\ell}^{H}$, can be modeled as \cite{Mingyao9693928}
 \begin{align} \label{channel:nf2}
\mathbf{h}_{\ell}^{H} = \sqrt{N} \beta_{\ell} \mathbf{a}_{\ell}^{H}( r_{\mathrm{u}}, \theta_{\mathrm{u}}),      \forall \ell\in \mathcal{L}.
 \end{align}
Herein,
$\beta_{\ell} = \frac{\lambda_{c}}{4\pi r_{\mathrm{u}}}$ is  the channel gain of subarray $\ell$,  $\mathbf{a}_{\ell}^{H}( r_{\mathrm{u}}, \theta_{\mathrm{u}})$ is its transmit steering vector,  $r_{\mathrm{u}}$ is the distance between the array center and the user, and   $\theta_{\mathrm{u}} \triangleq  \sin(\varphi_{\mathrm{u}})\in [-1, 1] $  is the spatial angle with $\varphi_{\mathrm{u}}$ being the angle-of-departure (AoD) from the BS array.  
{In particular, the $m$-th element of  $\mathbf{a}_{\ell}^{H}( r_{\mathrm{u}}, \theta_{\mathrm{u}})$, $[\mathbf{a}_{\ell}^{H}( r_{\mathrm{u}}, \theta_{\mathrm{u}})]_{m}$, is modeled as 
 \begin{align} \notag
[\mathbf{a}_{\ell}^{H}\!(  r_{\mathrm{u}}, \theta_{\mathrm{u}})]_{m} \!= \!\frac{1}{\sqrt{N}} e^{\frac{j2\pi r_{\mathrm{u}}}{\lambda_{c}}} e^{-j \frac{2\pi r^{((\ell-1)M+m-1)}}{\lambda_{c}}},
 \end{align} 
where  $r^{(n)}  = \sqrt{r_{\mathrm{u}}^{2} + \delta_{n}^{2} d^{2} - 2r_{\mathrm{u}}\theta_{\mathrm{u}} \delta_{n} d}$ denotes the distance between the $n$-th antenna and the user.}
Based on the Fresnel approximation \cite{Mingyao9693928}, $r^{(n)}$ can be approximated as\footnote{{
The channel model for the compact modular array in this paper can be further extended to the case of separated modular
arrays by setting the distance between the centers of adjacent subarrays to $Zd$, where $Z$ (\mbox{$Z\geq M$}) is the modular separation parameter.}} 
 \begin{align} \label{distancea:app}
r^{(n)} \approx  r_{\mathrm{u}}- \delta_{n}\theta_{\mathrm{u}} d + \frac{\delta_{n}^{2} d^{2}(1-\theta_{\mathrm{u}}^{2} )}{2r_{\mathrm{u}}},~\forall n\in \mathcal{N}. 
 \end{align}
{Note that \eqref{distancea:app}  is mathematically accurate when the distance between the user and the array is larger than $ 0.62 \sqrt{\frac{D^{3}}{\lambda_{c}}}$ \cite{7942128}.}

\subsection{Signal Model}
Let $x(t) \in \mathbb{C}$ denote the  complex-valued baseband signal sent by the BS  at  time slot $t$.  We consider the case where the time synchronization of data transmissions among different subarrays is imperfect due to hardware imperfections in practice. 
 Mathematically, the passband signal sent from  the $\ell$-th  subarray  can be expressed as
\begin{align}
\mathbf{x}_{\ell}(t) = \mathbf{w}_{\ell}x(t-\tau_{\ell}) e^{j2\pi f_{c}(t-\tau_{\ell})} ,  \forall \ell \in \mathcal{L},
 \label{TX:signal}
\end{align}
where 
 $\mathbf{w}_{\ell} \in \mathbb{C}^{M \times 1}$ is the beamforming vector applied to the  $\ell$-th  subarray, $\tau_{\ell}$ is the time synchronization error with respect to (w.r.t.) the common reference clock (with  $\tau_{\ell} = 0$ representing perfect time synchronization), and $f_{c}$ is the carrier frequency. In practice, the error $\tau_{\ell}$ can be on the order of 10 picoseconds (ps)  for fiber-based synchronization technology in mmWave  bands \cite{9994246}. 
After passing through the channel $\mathbf{h}_{\ell}^{H}$, the corresponding received passband signal is given by 
\begin{align}
{y}_{\ell}(t) = \re\{ \mathbf{h}_{\ell}^{H} \mathbf{w}_{\ell}x(t-\tau_{\ell}) e^{j2\pi f_{c}(t-\tau_{\ell})}\},~ \forall \ell \in \mathcal{L} .
 \label{bandpassSIG}
\end{align}
After down-converting the passband signal by multiplying it with the carrier $e^{-j2\pi f_{c}t}$, the received signal  in the baseband~is 
\begin{align}
\bar{{y}}_{\ell}(t) = &\mathbf{h}_{\ell}^{H} \mathbf{w}_{\ell}x(t-\tau_{\ell}) e^{-j2\pi f_{c}\tau_{\ell}}  \notag
\\
\overset{(a)} \approx &  \mathbf{h}_{\ell}^{H} \mathbf{w}_{\ell}x(t) {e^{-j\phi_{\ell}}}, ~ \forall \ell \in \mathcal{L}, 
 \label{lowbandpassSIG}
\end{align}
where  $\phi_{\ell} = 2\pi f_{c}\tau_{\ell}$ and  $(a)$ holds based on the approximation $x(t-\tau_{\ell}) \approx x(t)$. This approximation is valid when the time synchronization error is relatively small compared to the symbol duration, i.e., $\tau_{\ell} \ll 1/B$, where $B$ denotes the bandwidth. 
 Note that the time synchronization error $\tau_{\ell}$ induces an \emph{effective phase shift} to the received signal from subarray $\ell$. For example, for mmWave systems operating at $f_{c} = 30$~GHz, a time synchronization error of $|\tau_{\ell}| = 30$~ps   can cause a substantial phase shift of $5.65$ rad.

\section{Beam-Focusing Analysis under Time Synchronization Errors }\label{section:BFpatAna}
In this section, we characterize the near-field beam pattern of XL-arrays, when the subarrays suffer from time synchronization errors. Specifically, we show that for the two-subarray case, the beam-focusing effect still persists even in the presence of  time synchronization errors, while the focused location is deviated from the user location. Moreover, for the general case of multiple subarrays, the time synchronization errors incur a more severe \emph{imbricated} beam pattern, hence resulting in degraded communication performance.

To this end, we first define the synchronization-error-aware near-field beam pattern as follows.
\begin{definition} [Beam pattern under time synchronization errors] \emph{Consider time synchronization errors  $\{\tau_{\ell}\}_{\ell = 1}^{L}$ for the $L$ subarrays. The near-field beam pattern of  {beamformer $\mathbf{w}\triangleq [\mathbf{w}_{1}^{T},...,  \mathbf{w}_{L}^{T} ]^{T}$} is defined as 
the \emph{normalized} beam power at an arbitrary  location ${S}_{q} \triangleq (r_{q}, \theta_{q})$, which is mathematically expressed as
\begin{align} \label{BFFocus1}
f_{L}({S}_{q};\! \mathcal{V}) \!= \!\left| \sum_{\ell = 1}^{L}   e^{j \phi_{\ell} }  \mathbf{a}_{\ell}^{H}\!(r_{q}, \theta_{q})  \mathbf{w}_{\ell}  \right|,  \forall {S}_{q}\! \in \!\mathcal{S}_{\mathrm{NF}},
\end{align}
where   $\mathcal{V}\triangleq \{ \phi_{1}, ..., \phi_{L} \}$,  $\mathbf{a}_{\ell}^{H}(r_{q}, \theta_{q}) $ is the channel steering vector of an arbitrary near-field observation location ${S}_{q}$, {and $\mathcal{S}_{\mathrm{NF}} \triangleq \{r_{q}| r_{q}\leq 2 D^2/{\lambda_c}\}$ denotes the near-field region.} 
}
\end{definition}

For the beam-focusing analysis,  we consider  the MRT-based beamforming at the BS  under the premise of  perfect  time synchronization. As such, we have  $\mathbf{w}_{\ell} = \mathbf{a}_{\ell}(r_{\mathrm{u}}, \theta_{\mathrm{u}}), \forall \ell\in \mathcal{L}$. Hence, the  beam pattern in \eqref{BFFocus1} becomes 
\begin{align} \label{BFFocus2}
f_{L}({S}_{q};  \mathcal{V}) = \left| \sum_{\ell = 1}^{L}   e^{j \phi_{\ell} }  \mathbf{a}_{\ell}(r_{q}, \theta_{q})   \mathbf{a}_{\ell}^{H}(r_{\mathrm{u}}, \theta_{\mathrm{u}}) \right|.
\end{align}
To facilitate the  beam pattern  analysis and obtain useful insights, we first consider a typical case with  two subarrays (i.e., $L = 2$), and then extend the results to the general case with an arbitrary number of subarrays.

\vspace{-0.20in}
\subsection{Near-Field Beam Pattern of Two Subarrays ($L = 2$)}

For the two-subarray case, i.e., $L = 2$, its near-field beam pattern under time synchronization errors in \eqref{BFFocus2} reduces to
\begin{align} \label{BFFocus3}
f&_{2}({S}_{q};  \mathcal{V}) \notag
\\
 &= \frac{1}{N} \left|e^{j\phi_{1}}  \sum_{n = 0}^{M-1}e^{j\frac{2\pi}{\lambda}(r_{\mathrm{u}}^{(n)}- r_{q}^{(n)}) } + e^{j\phi_{2}}  \sum_{n = M}^{N-1}e^{j\frac{2\pi}{\lambda}(r_{\mathrm{u}}^{(n)}- r_{q}^{(n)}) } \right| \notag
  \\
   & \overset{(b)}\approx\frac{1}{N} \bigg|e^{j\phi_{1}} \sum_{n = -(N-1)/2}^{-1/2}\! \!\! \! e^{jn\pi(\theta_{q}- \theta_{\mathrm{u}})+ j \frac{2\pi}{\lambda}n^{2}d^{2}\left( \frac{1-\theta_{\mathrm{u}}^{2}}{2r_{\mathrm{u}}} -\frac{1-\theta_{q}^{2}}{2r_{q}} \right)} 
   \notag
  \\
  &~~~~+ e^{j\phi_{2}} \sum_{n = 1/2}^{(N-1)/2} e^{jn\pi(\theta_{q}- \theta_{\mathrm{u}})+ j \frac{2\pi}{\lambda}n^{2}d^{2}\left( \frac{1-\theta_{\mathrm{u}}^{2}}{2r_{\mathrm{u}}} -\frac{1-\theta_{q}^{2}}{2r_{q}} \right)}
\bigg|
   \notag
  \\
  & \triangleq \tilde{f}_{2}({S}_{q};  \mathcal{V}),~  \forall {S}_{q} \in \mathcal{S}_{\mathrm{NF}},
\end{align}
where $(b)$  holds  due to the distance approximation in \eqref{distancea:app}.  Note that 
different from the conventional time-synchronized case, the near-field beam pattern under time synchronization errors in \eqref{BFFocus3} is affected by  the \emph{effective} phase shifts (i.e., $\phi_{1}$ and $\phi_{2}$) and the spherical wavefront in a  coupled manner, making it difficult to obtain useful insights.

To tackle this issue, we first numerically plot in Fig. 2 the beam focused locations  of time-asynchronous subarrays over 50 random  synchronization errors $\{\tau_{1}, \tau_{2}\}$. These errors are independently drawn from $\mathcal{U}[-26, 26]$~ps, where $\mathcal{U}(\cdot)$ denotes the uniform distribution. Based on these results, an interesting and important observation is made on  the behavior of beam pattern at a \emph{user ring}, which is  defined below. 

\begin{definition} [User ring] \emph{Given a user location $(r_{\mathrm{u}}, \theta_{\mathrm{u}})$, the user ring is a ring that contains all positions (including user location itself) satisfying the following condition 
\vspace{-0.10in}
\begin{align} 
\mathcal{S}_{\mathrm{ring}} = \left\{  (r_{q}, \theta_{q}) \in \mathcal{S}_{\mathrm{NF}}   \bigg|\frac{1-\theta_{q}^{2}}{2r_{q}} = \frac{1-\theta_{\mathrm{u}}^{2}}{2r_{\mathrm{u}}}\right\}. 
\end{align}
} 
\end{definition} 

\begin{figure}
\captionsetup{singlelinecheck = false, justification=justified}
   \centering
   \includegraphics[width=2.5in]{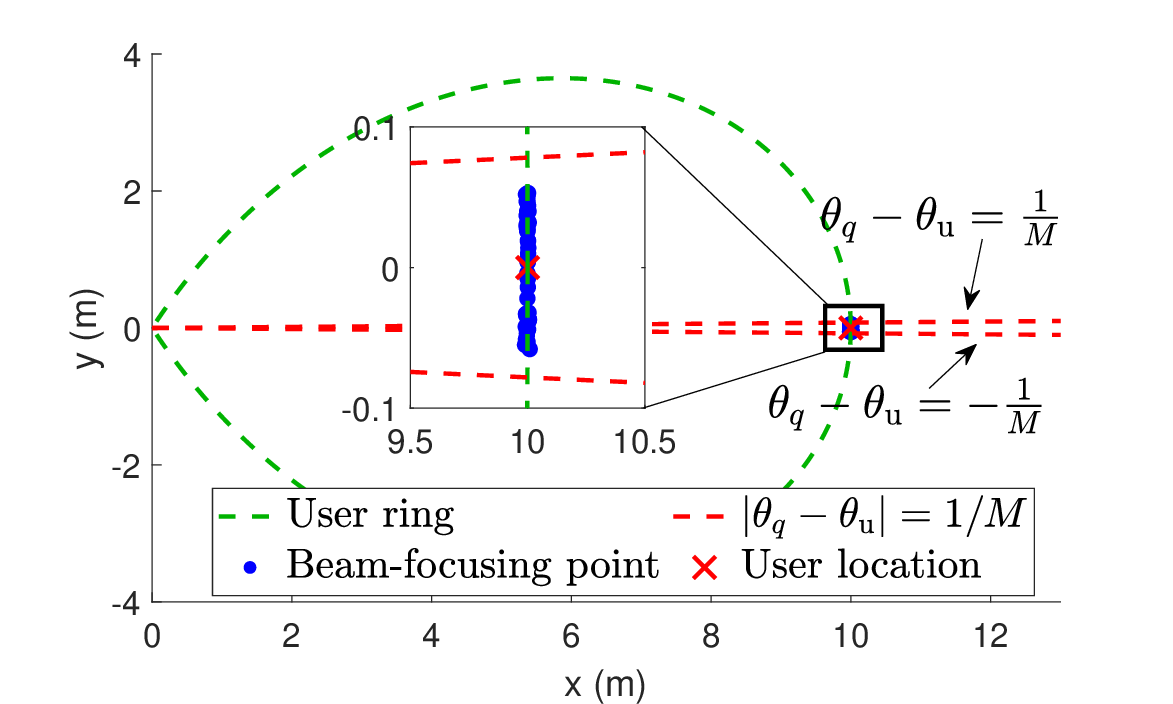}
   \vspace{-0.10in}
   \captionsetup{font={footnotesize}}
    \caption{Positions of the beam-focusing points (corresponding to the maximizers of $f_{2}({S}_{q}; \mathcal{V})$) given $L = 2$ and $M = 128$ over 50 random synchronization error realizations.}
   \vspace{-0.15in}
\label{disring}
\end{figure}

\noindent \textbf{\underline{\emph{Observation}} 1.}
In Fig. \ref{disring}, the beam-focusing points are located \emph{around} the user ring (as defined in \textbf{Definition 2}).\footnote {Obtaining precise  locations of beam-focusing points is challenging due to the intractability of $\tilde{f}_{2}({S}_{q};  \mathcal{V})$, which is thus left for future work.}

\textbf{Observation 1} shows that for the two-subarray case, the beam-focusing effect still \emph{persists} even in the presence of time synchronization errors, albeit with certain deviations in the spatial domain. {Moreover, for performance analysis, this observation allows us to focus on the beam pattern analysis at the user ring only, instead of the whole region. By exploiting mathematical properties of the user ring, the original beam pattern expression $\tilde{f}_{2}({S}_{q};  \mathcal{V})$, which involves an intractable two-dimensional distance and angle parameters, can be reduced to a simplified one-dimensional angular expression along the ring, as shown in the following lemma.}

\begin{lemma} \label{Lemma1} \emph{Consider the XL-array system with asynchronous transmissions at different subarrays. 
The beam pattern of $\mathbf{w}$ in \eqref{BFFocus3}, i.e.,  $\tilde{f}_{2}({S}_{q};  \mathcal{V})$,   at the  user ring can be re-expressed as
\begin{align} \label{BFFocus5}
\tilde{f}_{2}&({S}_{q};  \mathcal{V}) \notag
\\
 &=\frac{1}{N} \left|e^{j\phi_{1}} \!\! \! \! \sum_{n = -(N-1)/2}^{-1/2} \!\! \! \!  e^{jn\pi(\theta_{q}- \theta_{\mathrm{u}})} + e^{j\phi_{2}}  \!\! \! \!\sum_{n = 1/2}^{(N-1)/2} \!\! \! \! e^{jn\pi(\theta_{q}- \theta_{\mathrm{u}})} \notag
\right|
\\
&= \frac{2}{N} | g_{2, \mathrm{cos}}(\underline{\theta}; \mathcal{V})|\times  | g_{2, \mathrm{sin}}(\underline{\theta})|, ~\forall {S}_{q} \in \mathcal{S}_{\mathrm{ring}}, 
\end{align}
where $g_{2, \mathrm{cos}}(\underline{\theta}; \mathcal{V}) =  \cos\left( \frac{\phi_{1}-\phi_{2}   }{2} -  \frac{M\pi}{2}\underline{\theta} \right)$, $g_{2, \mathrm{sin}}(\underline{\theta})=  \frac{\sin(  {M\pi}\underline{\theta}/{2})}{\sin( {\pi} \underline{\theta}/{2} )}$, and $\underline{\theta} = \theta_q - \theta_{\mathrm{u}}$. 
}
\end{lemma}

\textbf{{Lemma \ref{Lemma1}}} shows that the phase-shift difference over the two subarrays arising from time asynchronization, i.e.,  ($\phi_{1}-\phi_{2}$),  appears in $g_{2, \mathrm{cos}}(\underline{\theta};\mathcal{V})$ only, which  captures  the impact of time synchronization errors,  whereas  $|g_{2, \mathrm{sin}}(\underline{\theta})| $ essentially governs the envelope of $\tilde{f}_{2}({S}_{q};  \mathcal{V}) $.  
Moreover, as shown in Fig.~\ref{Twosubarray:2D}, such a subarray-wise phase-shift difference causes a \emph{beam offset}, which deviates from the original target angle and/or distance. 
{
\begin{remark}[Extension to separated subarrays] \emph{The results in Lemma 1 can be extended to the case of separated subarrays with non-zero inter-subarray spacing. Generally speaking, the corresponding beam pattern properties are more complicated than the compact modular array due to the existence of grating lobes arising from non-zero subarray spacing \cite{Li10545312}, \cite{Gon11122484}. For the main lobe, it can be shown that its beam pattern has a similar form to (10) for the compact case. In particular, when the distance between the centers of adjacent subarrays is $Zd$, where $Z ~(Z \geq M)$ is the modular separation parameter,  the main-lobe beam pattern can be obtained by  substituting 
$M$ in $g_{2, \mathrm{cos}}(\underline{\theta};\mathcal{V})$ with $Z$. The property analysis of grating lobes is more complicated, which is left for future work.}
\end{remark}}

To mathematically characterize the effect of asynchronous transmissions over the two subarrays, we present the following Lemma.
\vspace{-0.10in}
\begin{lemma} \label{Lemma2} \emph{Given the beamformer $\mathbf{w}$ and synchronization errors $\{\tau_{1}, \tau_{2}\}$, the angle ${\theta}_{\mathrm{max}}$ that maximizes the beam power at the user ring $\mathcal{S}_{\mathrm{ring}}$ is given by}
\vspace{-0.05in}
  \begin{align} \label{Max:offset}
(r_{\mathrm{max}}, {\theta}_{\mathrm{max}} ) = \arg \max_{{S}_{q} \in \mathcal{S}_{\mathrm{ring}} } \tilde{f}_{2}({S}_{q};  \mathcal{V}). 
\end{align}
\end{lemma}
 \vspace{-0.05in}
The (beam-focusing) angle offset over the user ring between the original targeted angle $ \theta_{\mathrm{u}}$ and the true focused angle ${\theta}_{\mathrm{max}}$ can be obtained as $ \Delta\theta_{\mathrm{max}} = {\theta}_{\mathrm{max}} - \theta_{\mathrm{u}} $. Note that the explicit value of $\Delta\theta_{\mathrm{max}}$ generally can be numerically obtained by e.g., the exhaustive search. To obtain useful insights, we present a key property of $ \Delta\theta_{\mathrm{max}}$ as follows.

\vspace{-0.05in}
\begin{proposition} \label{Pro1} \emph{
For the two-subarray case with synchronization errors $\{\tau_{1},\tau_{2}\}$, the beam-focusing angle offset at the user ring is upper-bounded by $1/M$, i.e., $|\Delta\theta_{\mathrm{max}}| \leq 1/M$.
}
\end{proposition}

\begin{proof}
 {In Lemma 1,  $\left|g_{2,\sin}(\underline{\theta})\right|$ is the envelope term and is independent of $\{\phi_1,\phi_2\}$. It achieves its maximum at $\underline{\theta}=0$ and decreases monotonically when $|\underline{\theta}|\le 2/M$. In contrast, $\left|g_{2,\cos}(\underline{\theta};\mathcal V)\right|$ is periodic in $\underline{\theta}$ with period $2/M$. 
Let $\theta_{\rm peak}$ denote the location of the peak of $\left|g_{2,\cos}(\underline{\theta};\mathcal V)\right|$ that is closest to zero. Then we have $|\theta_{\rm peak}|\le 1/M$.  
Fig. \ref{Twosubarray:subfigProof} depicts four distinct scenarios for $\theta_{\rm peak}$: $\theta_{\rm peak} = 0$; $|\theta_{\rm peak}| = 1/M$; $0<\theta_{\rm peak}< 1/M$; and $-1/M<\theta_{\rm peak} <0$.
If $\theta_{\rm peak}=0$, then $\tilde f_2(S_q;\mathcal V)$ achieves its maximum at $\underline{\theta}=0$, which gives $\Delta\theta_{\max}=0$. Otherwise, we first consider the case $0<\theta_{\rm peak}\le 1/M$; while the proof for the case of $-1/M\le \theta_{\rm peak}<0$ follows similar procedures. Specifically, in the interval $[0,\theta_{\rm peak}]$, $\left|g_{2,\sin}(\underline{\theta})\right|$ decreases monotonically, while $\left|g_{2,\cos}(\underline{\theta};\mathcal V)\right|$ increases monotonically. On the other hand, in the interval $[\theta_{\rm peak},1/M]$, both terms decrease monotonically. Therefore,  $\tilde f_2(S_q;\mathcal V)$ achieves its maximum value in $[0,\theta_{\rm peak}]$, i.e.,  $0\le \Delta\theta_{\max}\le \theta_{\rm peak}\le 1/M$ (Note that $\Delta\theta_{\max}=\theta_{\max}-\theta_{\mathrm{u}}$).
Similarly, when $-1/M\le \theta_{\rm peak}<0$, we have $-1/M\le \theta_{\rm peak}\le \Delta\theta_{\max}\le 0$.
Combining the above yields $|\Delta\theta_{\max}|\le 1/M$,  thus completing the proof.}
\end{proof}

\noindent
\textbf{\underline{\emph{{Example}}} 1.}~  {We consider a two-subarray case (i.e., $L = 2$) with the user located at $(10, 0)$ meters (m) and $N = LM = 2\times128$. The beam-focusing points for 50 random realizations of $\{\tau_{1}, \tau_{2}\}$ are illustrated in Fig. \ref{disring}. Here, $\tau_{\ell} \sim\mathcal{U} [-26, 26]$~ps, $\forall \ell \in \{1,2\}$. It is observed from Fig. \ref{disring} that these beam-focusing points are distributed in the vicinity of the user ring, and their (spatial) angles fall within the interval $[-1/M, 1/M]$.

\begin{figure}
\captionsetup{singlelinecheck = false, justification=justified}
   \centering
   \includegraphics[width=2.3in]{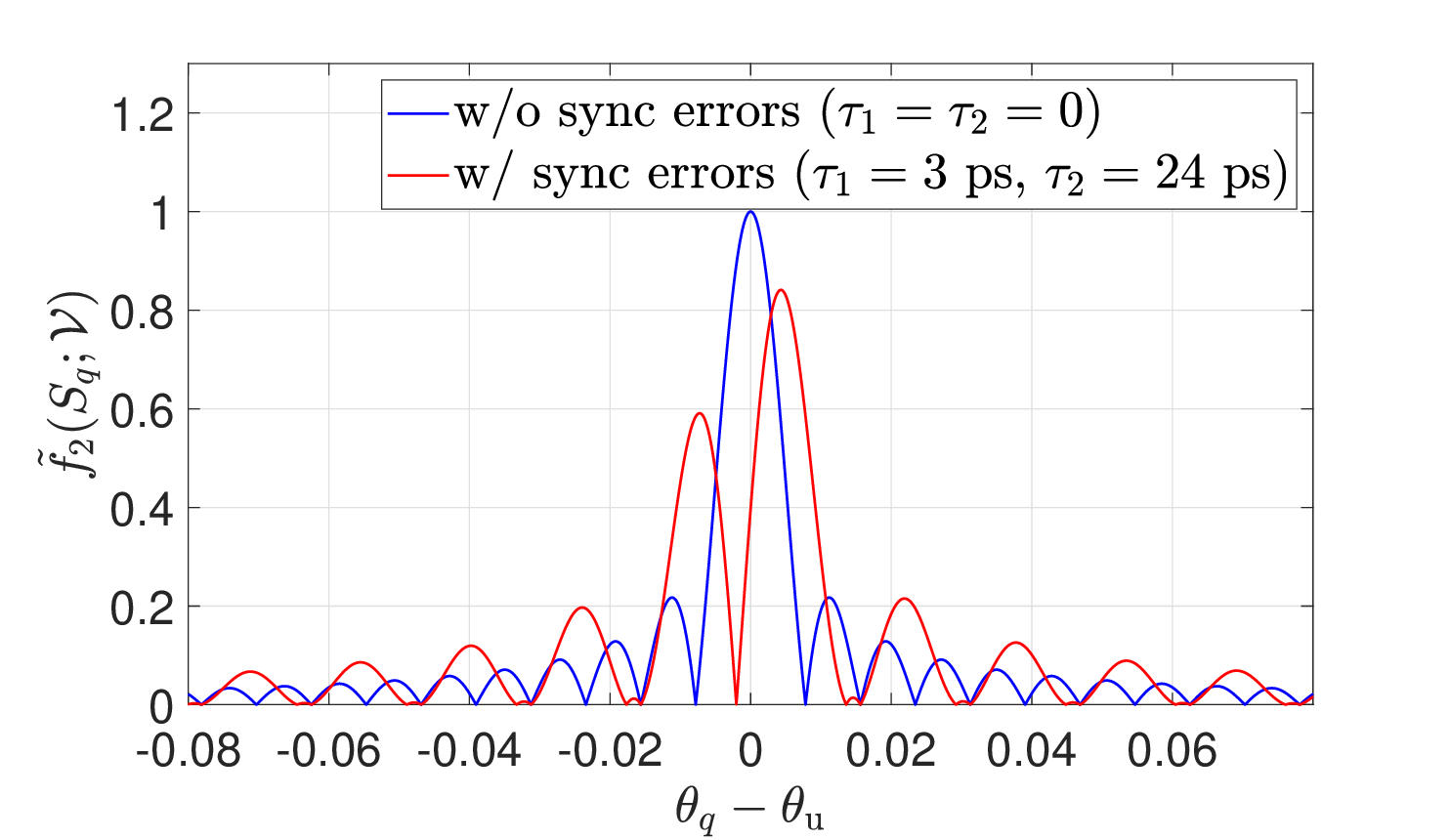} 
   \vspace{-0.10in}
   \captionsetup{font={footnotesize}}
    \caption{Beam patterns of two subarrays at the user ring $\mathcal{S}_{\mathrm{ring}}$.}
    \vspace{-0.25in}
\label{Twosubarray:2D}
\end{figure}


\begin{figure}
\captionsetup{singlelinecheck = false, justification=justified}
   \centering
   \includegraphics[width=3.0in]{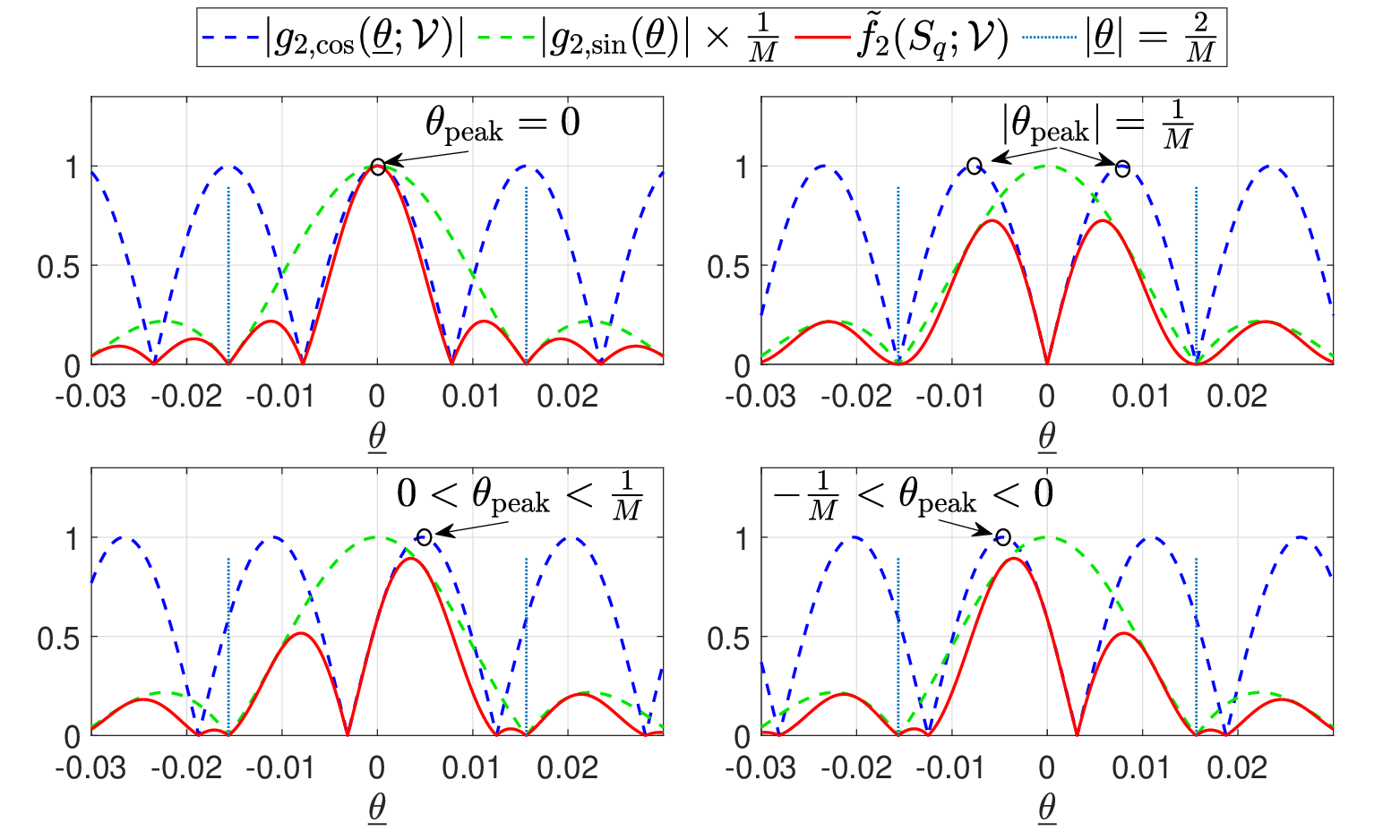} 
   \captionsetup{font={footnotesize}}
    \caption{Illustration of $\tilde{f}_{2}({S}_{q}; \mathcal{V})$, $| g_{2, \mathrm{cos}}(\underline{\theta};\mathcal{V})|$, and $ | g_{2, \mathrm{sin}}(\underline{\theta})|$ for different $\theta_{\mathrm{peak}}$.}
    \vspace{-0.25in}
\label{Twosubarray:subfigProof}
\end{figure}

\begin{remark} \emph{\textbf{{Proposition \ref{Pro1}}} shows that for the two-subarray case, regardless of the magnitude of the effective phase-shift difference, the beam-focusing angle shift at the user ring is always \emph{upper-bounded} by $1/M$, which decreases with the number of antennas per  subarray. {This beam-focusing offset not only reduces the received signal power at the target location/region, but also increases undesired interferences to  nearby users.} Therefore, to improve the communication performance, it is essential to achieve high-precision time synchronization (e.g., on the order of ps or even femtoseconds (fs) in mmWave bands) for subarrays. However, this is also more challenging for XL-arrays, as maintaining such synchronization accuracy among subarrays generally necessitates high-cost hardware and complex system design \cite{Tian11359250}.
}
\end{remark}

 \vspace{-0.25in}
\subsection{Near-Field Beam Pattern of Multiple Subarrays} \label{subsec:multisub}
Next, we consider the more general case with multiple subarrays (i.e., $L>2$). For this case, the beam pattern with time synchronization errors can be approximated as 
\begin{align} \label{BFFocus:Multi}
{f}&_{L}({S}_{q}; \mathcal{V})
 = \left| \sum_{\ell = 1}^{L}   e^{j\phi_{\ell}}\mathbf{a}_{\ell}^{H}(r_{\mathrm{u}}, \theta_{\mathrm{u}}) \mathbf{a}_{\ell}(r_{q}, \theta_{q}) \right| \notag
 \\
 & \overset{(c)} \approx \frac{1}{N}  \left| \sum_{\ell = 1}^{L}  \sum_{m = 0}^{M-1}   e^{j\phi_{\ell}}   e^{jm_{\ell}\pi(\theta_{q}- \theta_{\mathrm{u}})+ j \frac{2\pi}{\lambda}m_{\ell}^{2}d^{2}\left( \frac{1-\theta_{\mathrm{u}}^{2}}{2r_{\mathrm{u}}} -\frac{1-\theta_{q}^{2}}{2r_{q}} \right)}    \right|  \notag
  \\
 & \triangleq \tilde{f}_{L}({S}_{q}; \mathcal{V}), ~ \forall {S}_{q}\! \in \!\mathcal{S}_{\mathrm{NF}},
\end{align}
{where $(c)$ holds  according to the distance approximation in \eqref{distancea:app}, and $m_{\ell} = (\ell-1)M+m-(N-1)/2, \forall \ell \in \mathcal{L}$.}

{Note that compared with the two-subarray case, the near-field beam pattern analysis for the general case is more involved due to the quadratic phase term $\frac{2\pi}{\lambda}m_{\ell}^{2}d^{2}\left( \frac{1-\theta_{\mathrm{u}}^{2}}{2r_{\mathrm{u}}} -\frac{1-\theta_{q}^{2}}{2r_{q}} \right)$, which makes it highly difficult  to analytically characterize the properties of the resulting beam pattern.} To obtain useful insights, we plot in Fig. \ref{MultiBeam} the beam patterns of multiple subarrays, where the subarray synchronization errors $\{\tau_{\ell}\}_{\ell = 1}^{L}$ are distributed as $\tau_{\ell} \sim \mathcal{U} [-26, 26]$~{ps}, $\forall \ell \in \mathcal{L}$. Several key observations are summarized as follows.
\begin{itemize}
    \item \textbf{(Beam split)} For the case with multiple subarrays,  the appealing single beam-focusing effect \emph{disappears} in general when  time synchronization errors are significant (e.g., 26 ps). {Instead, the resulting beam splits into multiple ones (not all located around the user ring in general), exhibiting a new \emph{imbricated} beam pattern. This phenomenon arises because the random phase shifts introduced by time synchronization errors prevent the signals from coherently superimposing at the user's location. 
        Instead, at certain unintended spatial locations, the signals from a subset of subarrays may coincidentally align in phase, forming localized coherent superpositions. As a result, the emergence of these multiple unintended energy peaks exhibits an imbricated beam pattern.} This observation indicates that for a modest/large number of subarrays, their time synchronization in data transmissions is highly important, which, if not properly addressed, may incur severe inter-user interference and power loss.
    \item \textbf{(Beam offset angle)} {Similar to the two-subarray case, it is observed that, the (dominant) imbricated beam pattern induced by time synchronization {errors} is primarily confined within the angular region $[-\frac{1}{M}, \frac{1}{M}]$. This provides an efficient estimation for the beam offset region, which determines the interference zone caused by  timing offsets.} Moreover, as the number of subarrays increases (and consequently fewer antennas per subarray), the imbricated beam pattern becomes more dispersed.
       This indicates that it is desirable to construct XL-arrays with a small number of subarrays (albeit of a higher hardware/fabrication cost), which is more robust to time synchronization errors.
\end{itemize}

 \vspace{-0.15in}
 \section {{Numerical Results}}
In this section, we numerically show the effect of time synchronization errors on the multi-user communication performance. 
We consider a scenario involving three single-antenna users located at the Cartesian coordinates $(10, 0)$ m, $(12, -0.2)$ m, and $(14, 0.4)$ m, respectively. Let $\mathbf{h}_{\ell,k}$ denote the channel from subarray $\ell$ to user $k$, and $\mathbf{w}_{\ell,k}$ denote the beamforming vector employed at subarray $\ell$ for user $k$, $\forall \ell \in \mathcal{L}$, $\forall k \in \mathcal{K}\triangleq\{1, 2, 3\}$.
The achievable rate of user $k$ in bits/second/Hertz (bps/Hz), denoted by $R_{k}$, is given by $R_{k} = \log_{2}(1\!+ \! \mathrm{SINR}_{k})$, where the signal-to-interference-plus-noise ratio (SINR) is obtained as 
 \vspace{-0.05in}
\begin{align} \notag
\mathrm{SINR}_{k} = \frac{P_{k}|\sum_{\ell = 1}^{L}\mathbf{h}_{\ell,k}^{H} \mathbf{w}_{\ell,k} {e^{-j\phi_{\ell}}}|^{2}}{\sum_{i = 1,i\neq k}^{3}P_{i}| \sum_{\ell = 1}^{L}\mathbf{h}_{\ell,k}^{H} \mathbf{w}_{\ell,i} {e^{-j\phi_{\ell}}}|^{2}+ \sigma_{k}^{2}},
\end{align}
with $P_{k}$ and $\sigma_{k}^{2}$ representing the BS transmit power and noise power at  user $k$, respectively. To evaluate the beam-focusing performance, we consider  the MRT scheme, i.e.,  $\mathbf{w}_{\ell,k}  = \frac{\mathbf{h}_{\ell,k}}{\sqrt{L}|\mathbf{h}_{\ell,k}|} $, $\forall \ell \in \mathcal{L},  \forall k \in \mathcal{K}$.   

In Fig. \ref{Rate:L}, we plot the sum-rate versus different synchronization error bounds, with system parameters set as $N = 256$, $P_{k}= 3.3$ mW, and $\sigma_{k}^{2} = -80$ dBm, $\forall k \in \mathcal{K}$. The synchronization error $\tau_{\ell} \sim \mathcal{U}[-T, T]$ ps, $\forall {\ell} \in \mathcal{L}$. The key observations are summarized as follows. First,  when  $T$ increases, the sum-rate first decreases sharply and then fluctuates when $T$ is larger than a threshold (around 15 ps). 
This is intuitively expected since the phase-shift term $e^{j\phi_{\ell}}$ is a periodic function w.r.t. $\tau_{\ell}$. Second, the rate loss becomes more severe for a larger $L$. This degradation is attributed to two main factors caused by time synchronization errors: 1) the beam-split effect (as shown in Fig. \ref{MultiBeam}), which reduces the received signal power; and 2) more severe inter-user interference, which is consistent with the  results of Proposition 1.

\begin{figure}[!t]
    \centering
    \captionsetup[subfloat]{
        font=footnotesize,
        labelfont=rm,
        format=plain,
    }
    \subfloat[$L = 4$, $M = 64$.]{\includegraphics[width=0.51\columnwidth]{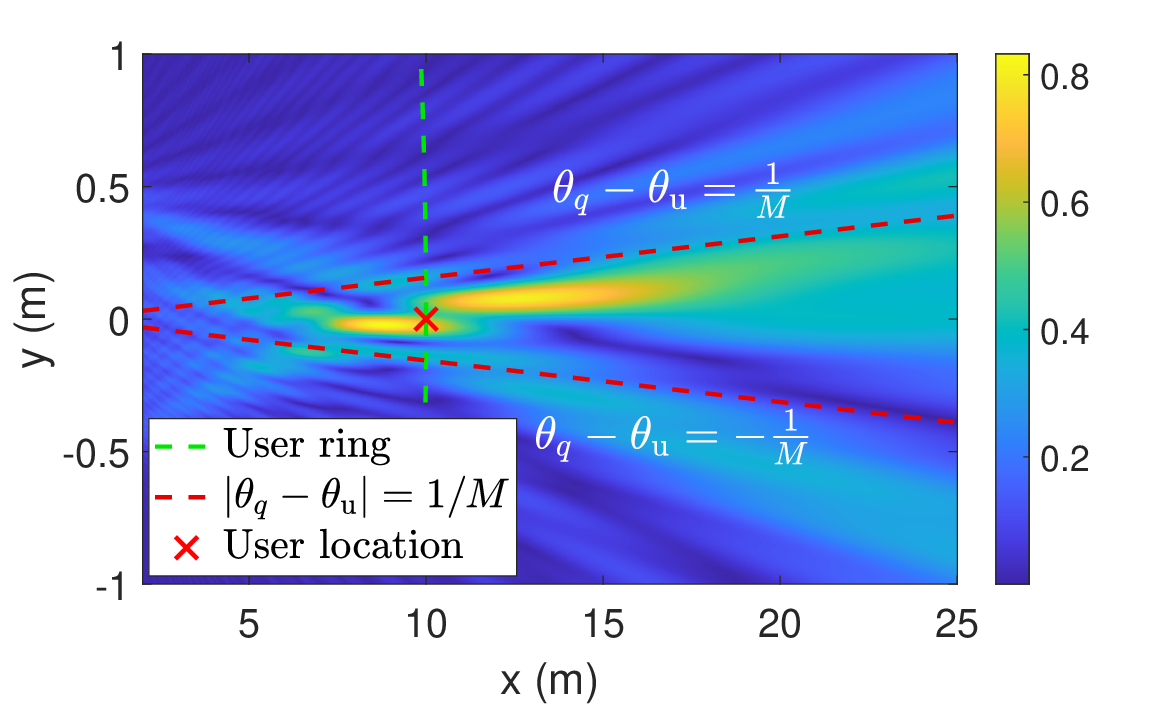}\label{fig_first_case}}%
    \hspace{-0.128in}%
    \subfloat[$L = 16$, $M = 16$.]{\includegraphics[width=0.51\columnwidth]{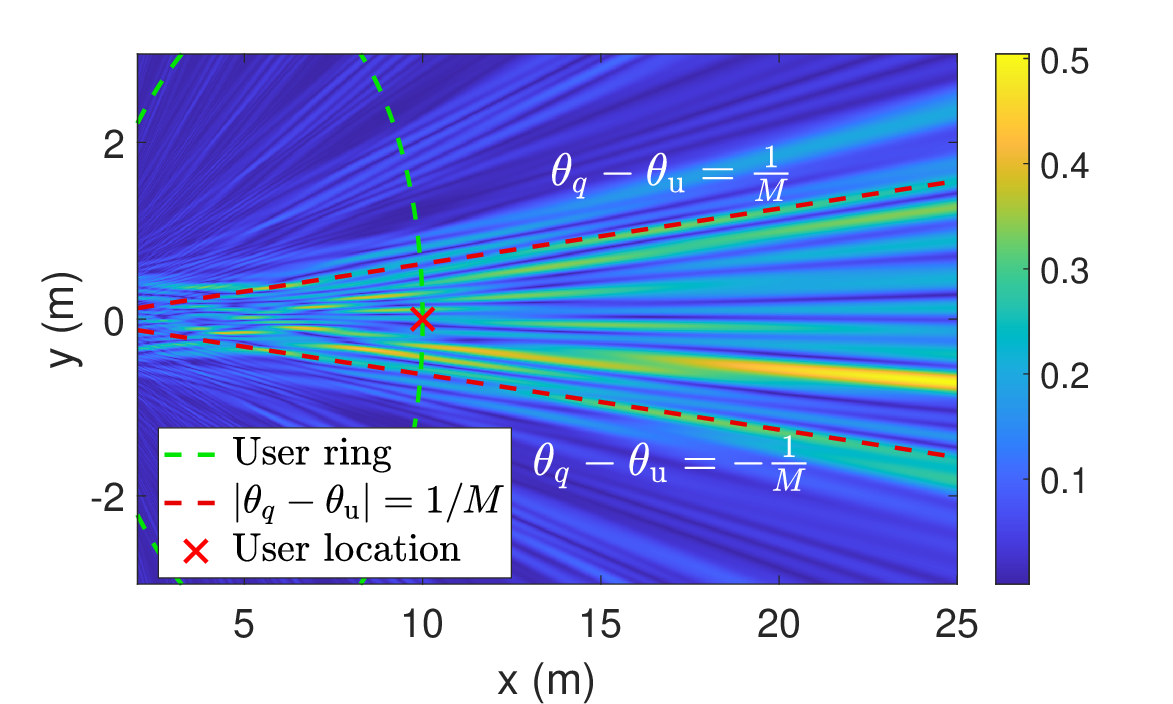}\label{fig_second_case}}
    \caption{Beam pattern with different $L$ and $M$ for a user located at (10, 0)~m.}
    \vspace{-0.15in}
    \label{MultiBeam}
\end{figure}


\vspace{-0.10in}
\section{Conclusions}

In this letter, we investigated the effects of time synchronization errors on near-field beam pattern under a modular architecture. For the two-subarray case, we examined the beam behavior along the user ring and obtained analytical expression for the resulting beam pattern. Both analytical and numerical results demonstrated that time synchronization errors across subarrays induce a beam offset, with angle deviation confined within a given angular support. 
 Moreover, for the multi-subarray case, we  showed that time synchronization errors trigger a beam-split effect, leading to an imbricated beam pattern. 
 The numerical results on sum-rate showed that time synchronization errors in subarray structures  may cause significant performance degradation in practice. 

\vspace{-0.15in}

\begin{figure}
\vspace{-0.05in}
\captionsetup{singlelinecheck = false, justification=justified}
   \centering
   \includegraphics[width=2.2in]{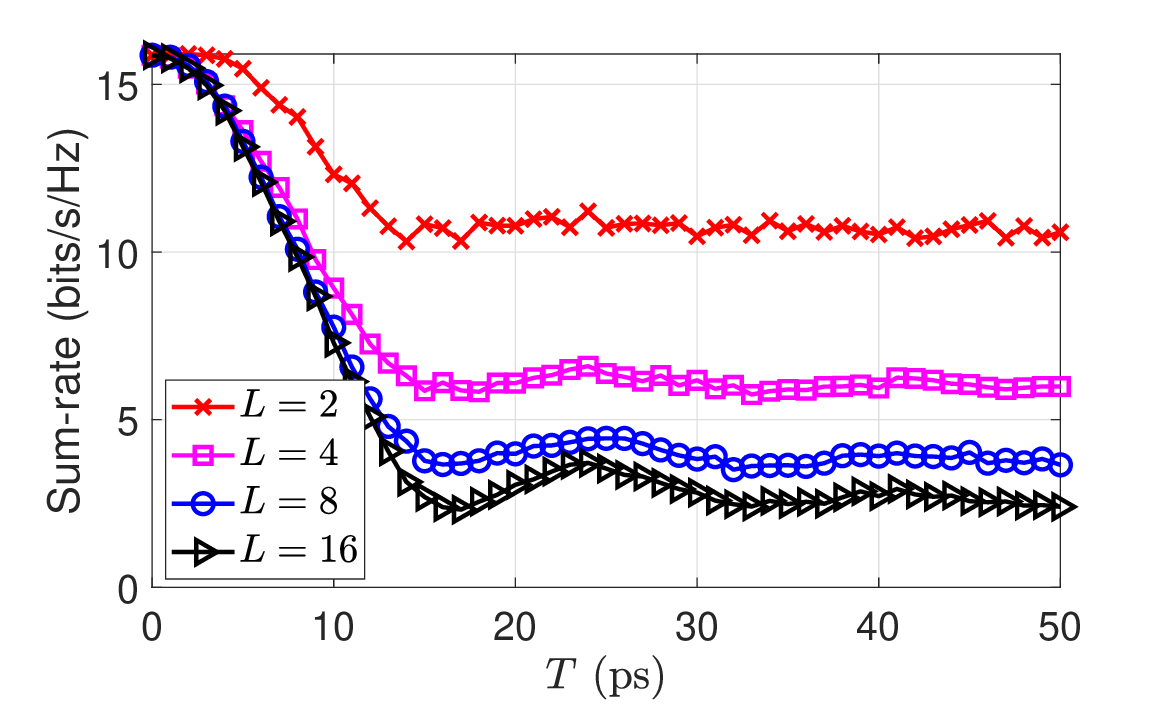}  
   \vspace{-0.10in}
   \captionsetup{font={footnotesize}}
    \caption{Sum-rate versus $T$.}
    \vspace{-0.1in}
\label{Rate:L}
\end{figure}


\footnotesize
\bibliography{IEEEabrv,ref_XLMIMOv2}
\bibliographystyle{IEEEtran}

%
\end{document}

%% file: header.tex
\newtheorem{definition}{\emph{\underline{Definition}}}

\newtheorem{lemma}{\emph{\underline{Lemma}}}

\newtheorem{proposition}{\emph{\underline{Proposition}}}

\newtheorem{remark}{\bf \emph{\underline{Remark}}}

\def\({\left(}
\def\){\right)}

\def\b0{{\mathbf{0}}}








%% file: main.bbl
\begin{thebibliography}{10}
\providecommand{\url}[1]{#1}
\csname url@samestyle\endcsname
\providecommand{\newblock}{\relax}
\providecommand{\bibinfo}[2]{#2}
\providecommand{\BIBentrySTDinterwordspacing}{\spaceskip=0pt\relax}
\providecommand{\BIBentryALTinterwordstretchfactor}{4}
\providecommand{\BIBentryALTinterwordspacing}{\spaceskip=\fontdimen2\font plus
\BIBentryALTinterwordstretchfactor\fontdimen3\font minus
  \fontdimen4\font\relax}
\providecommand{\BIBforeignlanguage}[2]{{%
\expandafter\ifx\csname l@#1\endcsname\relax
\typeout{** WARNING: IEEEtran.bst: No hyphenation pattern has been}%
\typeout{** loaded for the language `#1'. Using the pattern for}%
\typeout{** the default language instead.}%
\else
\language=\csname l@#1\endcsname
\fi
#2}}
\providecommand{\BIBdecl}{\relax}
\BIBdecl

\bibitem{You10858129}
C.~You \emph{et~al.}, ``Next generation advanced transceiver technologies for
  {6G} and beyond,'' \emph{IEEE J. Sel. Areas Commun.}, vol.~43, no.~3, pp.
  582--627, Mar. 2025.

\bibitem{Zhang10500404}
Y.~Zhang and C.~You, ``{SWIPT} in mixed near- and far-field channels: Joint
  beam scheduling and power allocation,'' \emph{IEEE J. Sel. Areas Commun.},
  vol.~42, no.~6, pp. 1583--1597, Jun. 2024.

\bibitem{Haiquan10496996}
H.~Lu \emph{et~al.}, ``A tutorial on near-field {XL-MIMO} communications toward
  {6G},'' \emph{{IEEE Commun. Surveys Tuts.}}, vol.~26, no.~4, pp. 2213--2257,
  Fourthquarter 2024.

\bibitem{Mingyao9693928}
M.~Cui and L.~Dai, ``Channel estimation for extremely large-scale {MIMO}:
  Far-field or near-field?'' \emph{IEEE Trans. Commun.}, vol.~70, no.~4, pp.
  2663--2677, Apr. 2022.

\bibitem{Haiyang9738442}
H.~Zhang \emph{et~al.}, ``Beam focusing for near-field multiuser {MIMO}
  communications,'' \emph{IEEE Trans. Wireless Commun.}, vol.~21, no.~9, pp.
  7476--7490, Sep. 2022.

\bibitem{Li10545312}
X.~Li, Z.~Dong, Y.~Zeng, S.~Jin, and R.~Zhang, ``Multi-user modular {XL-MIMO}
  communications: Near-field beam focusing pattern and user grouping,''
  \emph{IEEE Trans. Wireless Commun.}, vol.~23, no.~10, pp. 13\,766--13\,781,
  Oct. 2024.

\bibitem{Gon11122484}
J.~P. González-Coma, S.~Fernández, and F.~Javier López-Martínez, ``User
  selection in near-field gigantic {MIMO} systems with modular arrays,''
  \emph{IEEE Trans. Commun.}, vol.~73, no.~12, pp. 13\,357--13\,368, Dec. 2025.

\bibitem{Zhang11127210}
Y.~Zhang \emph{et~al.}, ``Modular {XL}-array-enabled channel estimation and
  localization in terahertz systems,'' \emph{IEEE Trans. Cogn. Commun. Netw.},
  vol.~12, pp. 3378--3392, 2026.

\bibitem{Tian11359250}
J.~Tian \emph{et~al.}, ``Pioneering scalable prototype for mid-band {XL-MIMO}
  systems: Design and implementation,'' \emph{IEEE J. Sel. Areas Commun.},
  Early access, Jan. 2026, doi: 10.1109/JSAC.2026.3656471.

\bibitem{Zhang11162269}
X.~Zhang, J.~Zhu, H.~Dai, W.~Huang, and W.~Liu, ``Joint synchronization and
  channel estimation in near-field distributed arrays,'' in \emph{Proc. IEEE
  Int. Conf. Commun. Wkshps (ICC Wkshps)}, Jun. 2025, pp. 757--762.

\bibitem{Zhang9913211}
Y.~Zhang, X.~Wu, and C.~You, ``Fast near-field beam training for extremely
  large-scale array,'' \emph{IEEE Wireless Commun. Lett.}, vol.~11, no.~12, pp.
  2625--2629, Dec. 2022.

\bibitem{7942128}
K.~T. Selvan and R.~Janaswamy, ``Fraunhofer and fresnel distances: Unified
  derivation for aperture antennas,'' \emph{IEEE Antennas Propag. Mag.},
  vol.~59, no.~4, pp. 12--15, Aug. 2017.

\bibitem{9994246}
J.~M. Merlo, S.~R. Mghabghab, and J.~A. Nanzer, ``Wireless picosecond time
  synchronization for distributed antenna arrays,'' \emph{IEEE Trans. Microw.
  Theory Techn.}, vol.~71, no.~4, pp. 1720--1731, Apr. 2023.

\end{thebibliography}
